\documentclass[10pt,aps,prl,twocolumn,superscriptaddress]{revtex4-2}
\usepackage[linkcolor = blue, citecolor = blue, urlcolor = blue, colorlinks = true]{hyperref}
\usepackage[usenames,dvipsnames,x11names]{xcolor}
\usepackage[normalem]{ulem}
\usepackage{graphicx}
\usepackage{listings}
\usepackage{stmaryrd}
\usepackage{amssymb}
\usepackage{amsmath}
\usepackage{gensymb}
\usepackage{float}
\usepackage{bbold}
\usepackage{bm}

\hypersetup{
    colorlinks=true,
    linkcolor=blue,
    citecolor=red}

\begin{document}
\title{Collective migration and topological phase transitions in confluent epithelia}

\author{Leonardo Puggioni}
\affiliation{Instituut-Lorentz, Universiteit Leiden, P.O. Box 9506, 2300 RA Leiden, The Netherlands}
\author{Dimitrios Krommydas}
\affiliation{Department of Physics, University of California Santa Barbara, CA 93106, USA}
\affiliation{Instituut-Lorentz, Universiteit Leiden, P.O. Box 9506, 2300 RA Leiden, The Netherlands}
\author{Luca Giomi}
\email{giomi@lorentz.leidenuniv.nl}
\affiliation{Instituut-Lorentz, Universiteit Leiden, P.O. Box 9506, 2300 RA Leiden, The Netherlands}

\date{\today}

\begin{abstract}
Collective epithelial migration leverages on topological rearrangements of the intercellular junctions, which allow cells to intercalate without loosing confluency. 
{\em In silico} studies have provided a clear indication that this process could occur via a two-step phase transition, where a hierarchy of topological excitations progressively transforms an epithelial layer from a crystalline solid to an isotropic liquid, via an intermediate {\em hexatic} liquid crystal phase. Yet, the fundamental mechanism behind this process and its implications for collective cell behavior are presently unknown. In this article, we show that the onset of collective cell migration in cell-resolved models of epithelial layers takes place via an activity-driven melting transition, characterized by an exponentially-divergent correlation length across the solid/hexatic phase boundary. Using a combination of numerical simulations and Renormalization Group analysis, we show that the availability of topologically distinct rearrangements -- known as T1 and T2 processes -- and of a non-thermal route to melting, renders the transition significantly more versatile and tunable than in two-dimensional passive matter. Specifically, the relative frequency of T1 and T2 processes and of the ``bare'' stiffness of the cell layer affect the divergence of positional correlations within a well-defined spectrum of critical behaviors. Suppressing T1 processes, changes the nature of the transition by preventing collective migration in favor of a cellular analog of surface sublimation.
\end{abstract}

\maketitle

Collective migration in epithelia occurs via a process known as epithelial-to-mesenchymal transition: i.e. the program of transcriptional switches and morphological transformations that allows cells to transition from an epithelial phenotype, characterized by strong intercellular adhesion and a regular spatial arrangement, to mesenchymal state, where cells are loosely connected and highly motile~\cite{Kalluri:2009}. This transition, as well as its inverse, the mesenchymal-to-epithelial transition, plays a vital role in a variety of cellular processes and developmental stages: from embryonic morphogenesis~\cite{Mongera:2018} to wound healing~\cite{Brugues:2014}, tissue fibrosis and cancer progression~\cite{Zhang:2018}. Yet, the paradigm of tissues as simple two-state systems, able to switch from stationary and solid-like to migratory and liquid-like in a single step, has recently been challenged by experimental evidence from both {\em in vitro} and {\em in vivo studies}~\cite{Montel:2021}. This has led to the concept of a continuum of intermediate phenotypes, with epithelial and mesenchymal representing the two extrema. These intermediate states, in turn, display distinctive cellular characteristics, including invasiveness, drug-resistance and the ability to form metastases at distant organs, thereby contributing to cancer metastasis and relapse.

\begin{figure}[t!]
\centering
\includegraphics[width=\columnwidth]{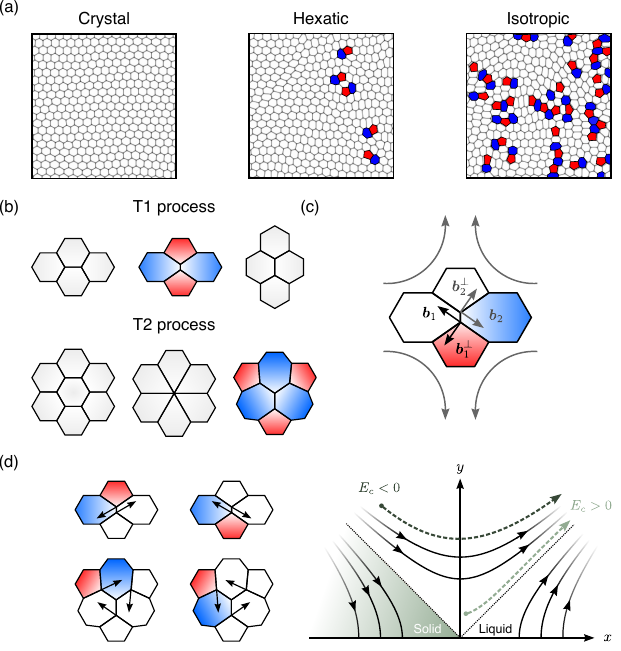}
\caption{\label{fig:1}(a) The three phases of epithelial layers from numerical simulations of the multiphase filed model. Upon increasing the speed of the cells, the system transitions from crystalline solid to hexatic liquid crystal and eventually to an isotropic liquid phase. Consistently with the KTNHY melting scenario~\cite{Nelson:1979,Young:1979,Kosterlitz:2016}, the first step of this transition is mediated by the unbinding of neutral pairs of $5$--$7$ dislocations and the second by the unbinding of the latter into isolated $5$- and $7$-fold disclinations. (b) Topological rearrangements process of the first (top) and second (bottom) kind, also known as T1 and T2 processes. (c) Schematic representation of the contribution of active stresses to the unbinding of dislocations. The gray arrows schematize the convergent-extension flow setting in response to a T1 processes. (d) Unbinding modes of complexes of $5$--$7$ dislocations in epithelia, with the black arrows marking the direction of the Burgers vector. (e) RG flow in the $xy$-plane. The magnitude of the exponent $\bar{\nu}$ depends on the material parameters via the initial condition $(x_{0},y_{0})$, here marked by a solid dot on the leftmost end of the dashed trajectories.} 	
\end{figure}

Because of its collective nature, several theoretical studies have attempted to rationalize the physical aspects of epithelial migration in the framework of phase transitions. Starting from the seminal work by Bi {\em et al}.~\cite{Bi:2015,Bi:2016}, this approach has progressively unveiled a picture that, while rich of model-dependent details, is consistent with a two-step process, typical of {\em passive} two-dimensional matter, known as Kosterlitz-Thouless-Halperin-Nelson-Young (KTHNY) melting scenario~\cite{Li:2018,Durand:2019,Pasupalak:2020,Li:2021,Li:2023,Nemati:2024}. The latter relates the increasing amount of disorder in a melting crystal to the proliferation of topological defects, known as {\em dislocations} and {\em disclinations}~\cite{Kosterlitz:1972,Kosterlitz:1974,Nelson:1979,Young:1979,Kosterlitz:2016} (see Fig.~\ref{fig:1}a). In the first step, the unbinding of neutral pairs of dislocations causes a reduction of {\em positional} order, but preserves a residual {\em orientational} order originating from the $p$-fold  symmetry (i.e. symmetry with respect to rotations by $2\pi/p$, with $p$ an integer) inherited from the lattice structure of the molten crystal. This results in the emergence of $p$-atic liquid crystal order, characterized by exponentially decaying (i.e. short-ranged) positional order and power-law decaying (i.e. quasi long-ranged) orientational order. A further increase in temperature, drives the transition to an isotropic liquid, where both translational and orientational order are short-ranged. The latter is mediated by the unbinding of dislocations into pairs of $\pm 1/p$ disclinations: i.e. singularities around which the average orientation of the building blocks rotates by $\pm 2\pi/p$. In the most common case of two-dimensional crystals with triangular lattice structure, $p=6$ and this scenario implies the existence of a {\em hexatic} phase, characterized by short-ranged positional order and quasi long-ranged $6$-fold orientational order, intermediate between crystalline solid and isotropic liquid~\cite{Nelson:1979,Young:1979,Kosterlitz:2016}. 

Now, whereas the evidence provided by {\em in silico} studies of epithelia has greatly contributed placing the physics of collective migration in the realm of critical phenomena ~\cite{Mueller:2019,Jain2024b,Li:2018,Durand:2019,Pasupalak:2020,Li:2021,Li:2023,Nemati:2024}, a consistent theoretical picture is still lacking and the questions outnumber the answers by far. Unlike colloidal particles, emulsion droplets, microgels etc., epithelial cells are mechanically active and organized in confluent layers, with no gaps and interstitial structure that cells can occupy after detaching from their neighbors. Furthermore, while critical phenomena are amenable to classification in terms of critical exponents, its is presently unclear to what extent the same program can be extended to the onset of collective epithelial migration and collective cellular processes in general. The latter severely limits the predictive power of current theories of collective cell migration, thus preventing their systematic experimental scrutiny. 

To overcome these limitations, in this article we propose a generalization of the classic KTHNY scenario to confluent epithelial layers. Our approach is build upon the assumption -- whose validity is verified {\em a posteriori} by means of numerical simulations -- that at criticality, where the effect of the active forces is comparable with that of fluctuations, the transition is effectively equilibrium-like. Taking advantage of the recently established correspondence between topological rearrangements in epithelia and hexatic defects~\cite{Krommydas:2024}, we show how active forces lead to a renormalization of the so-called defect {\em core energy}: i.e. the energetic cost associated with the breakdown of hexatic order at short distances from a defect. Confluency, on the other hand, enters in the melting process by reducing the degeneracy entailed with the unbinding of dislocations, thus affecting the specific structure of the Renormalization Group (RG) flow. 

Epithelial migration relies on a process known as {\em topological rearrangement process of the first kind}, or T1 for brevity, through which the vertices of a honeycomb  network merge and then split, thereby leading to a remodeling of the network’s topology (Fig.~\ref{fig:1}b and Refs.~\cite{Weaire:1999,Henkes2020,Beatrici2023,Claussen2024,Brauns2024}). In the lexicon of hexatic liquid crystals, this process is equivalent to the unbinding of a pair of $5$--$7$ dislocations with anti-parallel Burgers' vector: i.e. $\bm{b}_{1}=-\bm{b}_{2}$ (see Fig.~\ref{fig:1}c). Dislocations, in turn, are subject to Peach-Koehler forces when embedded in the stress field $\bm{\sigma}$ sourced by other dislocations or externally applied deformation: $F_{i}=\epsilon_{ij}\sigma_{jk}b_{k}$, with $\bm{\epsilon}$ the anti-symmetric tensor (i.e. $\epsilon_{xy}=-\epsilon_{yx}=1$ and $\epsilon_{xx}=\epsilon_{yy}=0$). These forces play a crucial role in solids, where they drive the motion of dislocations, ultimately resulting into plastic deformations. In epithelia, the stress field $\bm{\sigma}$ is instead internally generated by the cells at the onset of intercalation. This gives rise to an effective potential energy field -- i.e. $U_{\rm eff}=-\int_{\bm{r}}^{\bm{r}+\delta\bm{r}}{\rm d}\bm{r}'\cdot\bm{F}$ -- associated to the mechanical work performed by the Peach-Koehler forces to displace a dislocation a distance $\delta\bm{r}$ away from its initial position. To make progress, we model $\bm{\sigma}$ as the viscous stress originating from the shear motion of the unbinding dislocations: i.e. $\bm{\sigma}= \eta [\nabla\bm{v}+(\nabla\bm{v})^{\intercal}]$, where the shear viscosity $\eta$ here embodies the lateral interactions resulting from intercellular adhesion and $\bm{v}$ is the velocity field associated with the convergent-extension flow of the intercalating cells. At short distances from the core, $\bm{v} \approx \dot{\epsilon}(x_{1}\bm{e}_{1}-x_{2}\bm{e}_{2})$, with $\{\bm{e}_{1},\bm{e}_{2}\}$ the Cartesian frame centered at the core and oriented as illustrated in Fig.~\ref{fig:1}c, $(x_{1},x_{2})$ local coordinates and $\dot{\epsilon}$ a constant shear rate. This structure can be derived within the framework of active hexatic hydrodynamics (see Supplementary Sec.~S1), or assumed on the basis of an empirical knowledge of T1 processes. From the above, it follows that
\begin{equation}\label{eq:active_stress}
\bm{\sigma} = \sigma\,\left(\frac{\bm{b}\otimes\bm{b}^{\perp}+\bm{b}^{\perp}\otimes\bm{b}}{|\bm{b}|^{2}}\right)\;,	
\end{equation}
where $\sigma=2\eta\dot{\epsilon}$ is the stress magnitude, $\bm{b}=\bm{b}_{1}=-\bm{b}_{2}$ at the time of unbinding and the superscript $\perp$ denotes a counter-clockwise $90^{\circ}$ rotation. Eq.~\eqref{eq:active_stress} implies $\bm{F}=\sigma\bm{b}$ and $U_{\rm eff}=-\sigma\bm{b}\cdot\delta\bm{r}$. Now, in order to completely unbind, the dislocations must move away from each other by a distance $|\bm{b}|$. Taking $\delta\bm{r}=-\bm{b}$ in $U_{\rm eff}$ leads to the conclusion that active forces effectively contribute to the melting process by shifting the core energy of unbinding dislocations by an amount proportional to the active shear stress $\sigma$. That is
\begin{equation}\label{eq:core_energy}
E_{c} = \sum_{i}(\epsilon_{c}+\sigma)|\bm{b}_{i}|^{2}\;,	
\end{equation}
with $\epsilon_{c}$ the passive core energy per unit area~\cite{Nelson:1979,Young:1979}. We stress that both terms in Eq.~\eqref{eq:core_energy} are {\em potential} and vanish identically in the solid phase, where $\bm{b}_{i}=\bm{0}$. Intuitively, activity can either facilitate or impede the unbinding of dislocations, depending on whether extensile (i.e. $\sigma<0$) or contractile (i.e. $\sigma>0$). The former is consistent with experimental observations on migrating layers of Madin–Darby canine kidney (MDCK) cells~\cite{Balasubramaniam:2021}.  

After having accounted for the role of activity, we next investigate that of confluency. In the classic KTHNY scenario, where particles are treated as point-like, Nelson and Halperin identified six distinct unbinding modes involving neutral complexes of two and three $5$--$7$ dislocations respectively (Fig.~\ref{fig:1}d and Ref.~\cite{Nelson:1979}). Each of these mode, in turn, contributes to the renormalization of the material parameters of the system. In elastically isotropic materials, these are the Young modulus $Y$, the core energy $E_{c}$ and the temperature $T$, from which one can construct two independent scaling fields: i.e. the {\em softness} parameter $x=K^{-1}-K_{c}^{-1}$, with $K=Ya^{2}/(k_{B}T)$, $a$ the lattice spacing and $K_{c}$ marking the critical point; and the {\em fugacity} $y=e^{-(\epsilon_{c}+\sigma)a^{2}/(k_{B}T)}$. Following Refs.~\cite{Nelson:1979,Young:1979} one finds
\begin{subequations}\label{eq:rg}
\begin{gather}
\frac{{\rm d}x}{{\rm d}l} = c_{2}y^{2}\;,\\
\frac{{\rm d}y}{{\rm d}l} = c_{1}xy+c_{3}y^{2}\;, 	
\end{gather}
\end{subequations}
with $l$ the RG time-like variable, and $c_{1}$, $c_{2}$ and $c_{3}$ positive constants reflecting the specific unbinding modes. Before proceeding towards the solution of Eqs.~\eqref{eq:rg}, two comments are in order. 

First, while the role of temperature in tissues is still largely unexplored, with the majority of {\em in vitro} experiments being performed at physiological conditions, thermal-like fluctuations are routinely included in theoretical models of epithelial layer (see e.g. Refs.~\cite{Bi:2015,Bi:2016,Li:2018,Durand:2019,Pasupalak:2020,Li:2021,Li:2023,Nemati:2024}). In these models, Gaussian white noise is generally introduced via a Langevin or Monte Carlo thermostat and is therefore indistinguishable from the thermal noise characteristic of Gibbs ensembles. While applied to epithelia, the factor $k_{B}T$ in the definition of $x$ and $y$ should therefore be interpreted -- in the spirit of Einstein's kinetic theory -- as proportional to the variance of any thermal or non-thermal noise field contributing to the cells' Brownian motion. 

Second, the constants $c_{1}$, $c_{2}$ and $c_{3}$ express the statistical weight of configurations featuring one, two or three dislocations respectively. Crucially, while $c_{1}$ is always finite, $c_{2}$ and $c_{3}$ depends upon the specific dislocation unbinding modes at play. In the particle systems with a triangular lattice structure, these constants have been calculated in Refs.~\cite{Nelson:1979,Young:1979} using the vector Coulomb gas model of interacting dislocations (see Supplementary Sec.~S2). This yields $c_{1}=32\pi$, $c_{2} \approx 2597.84$ and $c_{3} \approx 38.93$. By contrast, in {\em closed} epithelial layers, where the rates of cell division and apoptosis are dwarfed by that of cell intercalation is faster than any other process~\cite{Sakar2022,Aigouy2010}, dislocation unbinding occurs exclusively via T1 processes thus reducing the degeneracy of the unbinding modes listed in Fig.~\ref{fig:1}f from six to two, corresponding to the two equivalent choices of pairing $5$- and $7$-fold disclinations to form a $5$--$7$ dislocation. In particular, the lack of modes involving three simultaneously unbinding dislocations implies $c_{3}=0$, thus Eqs.~\eqref{eq:rg} reduce to the RG equations of the Kosterlitz-Thouless (KT) transition~\cite{Kosterlitz:1972,Kosterlitz:1974}. 

\begin{figure*}[t!]
\centering
\includegraphics[width=\textwidth]{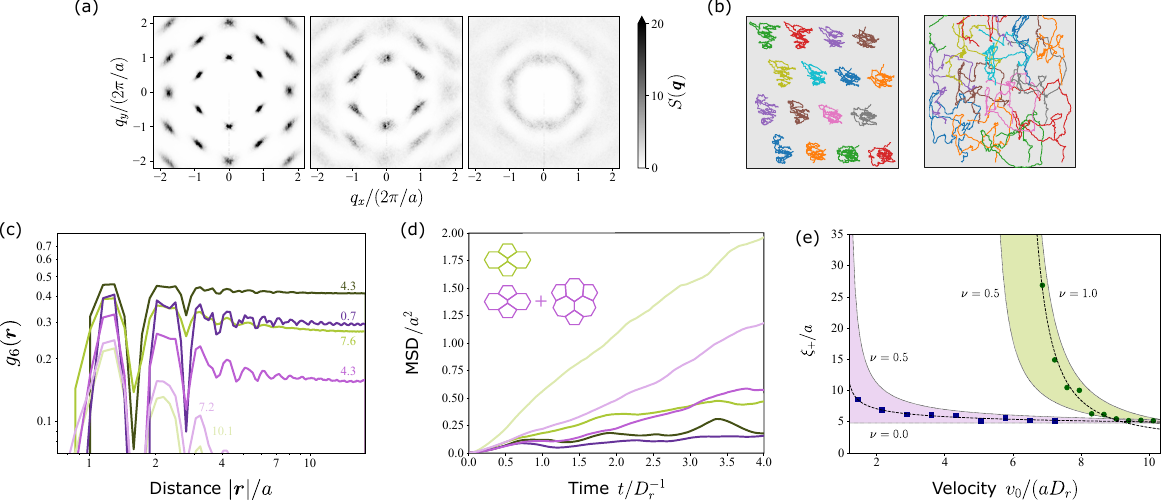}
\caption{\label{fig:2}Multiphase field simulations of the onset of collective migration in closed and open systems (a) Structure factor of closed systems in the crystalline, hexatic and isotropic liquid phase. (b) Trajectories of a selected number of cells across the melting transition (see Supplementary Fig.~S2 in Ref.~\cite{SI} for analogous results in open systems). (c) Orientational correlation function for varying self-propulsion speed $v_{0}$ in both closed (green tones) and open (magenta tones) systems. This function has a finite limit in the crystalline phase and decays as a power in the hexatic phase and exponentially in the isotropic phase. (d) Mean-squared displacement (MSD) in closed (green tones) and open (magenta tones) systems and the same $v_{0}$ values as in panel (c). (e) Correlation length $\xi_{+}$ as a function of the self-propulsion speed.} 	
\end{figure*}

A even richer scenario is that found in {\em open} epithelial layers, wherein cells can enter and exit by means of {\em topological rearrangement processes of the second kind}, also known as T2 processes (Fig.~\ref{fig:1}b and Ref.~\cite{Weaire:1999}). As vacancies and interstitial in crystals, these threefold features are analogous to neutral triplets of $5$--$7$ dislocations~\cite{Bowick:2007}, thus allow for finite $c_{3}$ values in Eq.~\eqref{eq:rg}. Following Refs.~\cite{Kosterlitz:1974,Nelson:1979,Young:1979,Kosterlitz:2016}, one can now use Eqs.~\eqref{eq:rg} to demonstrate that, when approached from the hexatic phase, the transition is marked by an exponentially divergent correlation length of the form
\begin{equation}\label{eq:xi}
\log\left(\xi_{+}/a\right) \sim |\sigma-\sigma_{c}|^{-\bar{\nu}}\;,
\end{equation}
where $\sigma_{c}$ is the critical stress and $\bar{\nu}$ a critical exponent. For $c_{3}=0$, Kosterlitz' seminal calculation yields $\bar{\nu}=1/2$~\cite{Kosterlitz:1974}. Remarkably, as we detail in Supplementary Sec.~S2, this result does not {\em always} carry over to epithelia, where, in the same regime, the exponent $\bar{\nu}$ spans the range $1/2\le \bar{\nu} \le 1$. More generally, when both T1 and T2 processes are available and $c_{3}>0$, we find
\begin{equation}\label{eq:nu_bar}
\frac{1}{2}\left(1-\frac{c_{3}}{\sqrt{c_{3}^{2}+4c_{1}c_{2}}}\right) \le \bar{\nu} \le 1\;.
\end{equation}
Because of the inherent nonlinearity of Eqs.~\eqref{eq:rg}, the rate at which the trajectories of the corresponding RG flow escape towards the liquid phase, corresponding to the limit $y\to\infty$ in Fig.~\ref{fig:1}e, depends on the material parameters, here embodied in the initial conditions $(x_{0},y_{0})$. In particular, it is possible to identify two distinct scaling regimes, depending on the distance between the trajectory emanating from the point $(x_{0},y_{0})$ and the dashed separatrices dividing $xy$-plane in Fig.~\ref{fig:1}e. When $E_{c}>0$, the initial fugacity $y_{0}$ is {\em exponentially small}, thus $(x_{0},y_{0})$ is close to the origin of the $xy$-plane and the associated trajectory remains in proximity of the separatrices along the RG flow. This behavior yields the lower bound of the range given in Eq.~\eqref{eq:nu_bar}, thus a number in the range $0\le\bar{\nu}\le 1/2$, depending on the ratio $c_{3}/\sqrt{c_{1}c_{2}}$, corresponding to the relative rate of T1 and T2 processes. When $E_{c}<0$, on the other hand, $y_{0}$ is {\em exponentially large} and the trajectory spans the region of the $xy$-plane where the critical exponent is close the upper bound $\bar{\nu} = 1$ (see Fig.~\ref{fig:1}e and Supplementary Sec.~S2 for details).
   
To test the significance of this analysis, we have numerically simulated the onset of collective cell migration in a multiphase field model (MPF) of confluent epithelia, whose significance in cell migration has been established by various recent studies~\cite{Nonomura:2012,Camley:2014,Palmieri:2015,Lober:2015,Marth:2016,Mueller:2019,Loewe2020,Hopkins:2022,Hopkins2023,Jain2024,Chiang2024a,Jain2024b,Chiang2024b}, in both closed and open systems. In our simulations, each cell represented by a two-dimensional scalar field $\phi_{i}=\phi_{i}(\bm{r},t)$, whose dynamics is determined by equilibrium and non-equilibrium processes via the equation
\begin{equation}\label{eq:mpf}
\partial_t \phi_i + \bm{v}_i \cdot \nabla \phi_i = - \frac{\delta \mathcal{F}}{\delta \phi_i}\;, \qquad i=1,\,2\dots\,N\;,
\end{equation}
where $\bm{v}_{i}$ is the velocity of the $i$-th cell and $\mathcal{F} = \mathcal{F}[\phi]$ a free-energy functional accounting for cells' preferential size, shape and mechanical interactions. These two quantities, in turn, are related by $\bm{v}_{i}=v_{0}\bm{p}_{i}+\zeta^{-1}\int {\rm d}A\,\phi_{i}\nabla(-\sum_{j=1}^{N}\delta\mathcal{F}/\delta\phi_{j})$, with $v_{0}$ a constant, $\bm{p}_{i}$ a unit vector subject to random rotations at rate $D_r$ and $\zeta$ a drag coefficient, resulting from the interactions with the substrate~\cite{Monfared2023,Monfared2024short} (see Supplementary Sec.~S3 for details). Compared to other cell-resolved models of epithelia, such as the Vertex and Voronoi models, the MPF model offers a more accurate description of cellular morphology and mechanical interactions, at the expense of a much higher computational cost~\cite{Alert:2020,Wenzel:2021,Moure:2021}. Now, in this setting, T1 processes occur spontaneously as a consequence of the cells' self-propulsion. By contrast, T2 processes can be achieved by opening the two-dimensional cell layer to external intrusion and extrusion events. The former are simply obtained by adding a new cell at a randomly selected vertex of the junction network. The latter, on the other hand, is here achieved upon boosting the compressibility of a randomly selected cell, so to make it collapse under pressure from the neighbors. The process is implemented in such a way to keep the total number of cells constant in average. Crucially, while the rate of T1 processes, associated with the coefficient $c_{2}$ in Eqs.~\eqref{eq:rg}, spontaneously emerges from the interplay between the active and passive forces at play -- with the former being sourced by self-propulsion and the latter by free-energy variations -- the rate of T2 processes, associated with the coefficient $c_{3}$ in Eqs.~\eqref{eq:rg}, is instead a control parameter of our numerical model.  

Our {\em in silico} epithelial layers, consists of $N=1085$ cells, whose centers are initially placed in a triangular lattice. After thermalization, we numerically calculate the system structure factor, $S(\bm{q})=\langle \rho_{\bm{q}}\rho_{-\bm{q}} \rangle/N$, with $\rho_{\bm{q}}$ the Fourier-transformed cell density and $\bm{q}$ the wave vector, and orientational correlation function $g_{6}(\bm{r})=\langle e^{6i[\theta(\bm{r})-\theta(\bm{0})]}\rangle$, with $\theta$ the orientation of the virtual ``bonds'' connecting the centers of the cells (the same conclusions can be drawn by considering the orientation of individual cells, see Supplementary Secs.~S3 and S4 and Refs.~\cite{Armengol2022a,Armengol2022b,Eckert2022}). To highlight the connection between hexatic order and collective migration, we further track the trajectories of the intercalating cells and compute their mean-squared displacement (MSD), as well as the densities of unbound defects, using the self-propulsion speed and the rate of intrusion and extrusions as control parameters. The latter, finally allows us to estimate the correlation length, by taking advantage of the fact that, near criticality, $n_{5\text{--}7} \sim \xi_{+}^{-2}$, with $n_{5\text{--}7}$ the number density if $5$--$7$ dislocations. Similar methods are used in experimental studies of melting in passive particle systems~\cite{Keim:2007}.    

In both closed and open system -- with the former characterized solely by T1 processes and the latter by both T1 and T2 processes -- increasing self-propulsion drives a two-step melting transition, with the solid and isotropic liquid phases separated by an intermediate hexatic phase. In the crystalline phase, the structure factor is characterized by finite-width Bragg peaks reflecting the presence of quasi long-ranged positional order~\cite{Nelson:1979}. After melting, these peaks elongate into segments, arising from the quasi long-ranged $6$-fold orientational order of the hexatic phase, to eventually merge into a ring once hexatic order becomes short-ranged (Fig.~\ref{fig:2}a). We stress that, even in the solid phase, the cells persistently random walk about their initial positions, but never undergo a complete intercalation (Fig.~\ref{fig:2}b). In this respect, solid epithelial layers may share similarities with the ``caged'' active solids experiementally investigated in Ref.~\cite{Baconnier:2022}. The same three phases can be also identified from the orientational correlation function, whose behavior at large distances transitions from constant to power-law decaying to exponentially decaying (Fig.~\ref{fig:2}c). Whereas the difference between closed and open systems does not shift the phase behavior away from the KTNHY scenario, this becomes prominent in the light of collective migration. 

In open systems, T2 processes produce local rearrangements of the cells even at small $v_{0}$ values, thus rendering the epithelial layer consistently more fluid, in agreement with the observation of Refs.~\cite{Bi2024,Silke2017}. Consequently, cells collectively migrate over a significantly broader range even if modestly motile (Fig.~\ref{fig:2}d). Measuring the correlation length confirms the softening caused by T2 processes, as well as the general picture provided by our RG analysis. Specifically, while Eq.~\eqref{eq:xi} in recovered both closed and open systems, with $\sigma \sim v_{0}$, fitting the exponent $\bar{\nu}$ gives $\bar{\nu} = 1.0 \pm 0.1$ in closed and $\bar{\nu} = 0.18 \pm 0.06$ in open systems (Fig.~\ref{fig:2}e). Thus, in closed systems, where melting is driven exclusively by activity in the range of elastic moduli considered here, $\bar{\nu}$ is at the upper bound of the range given by Eq.~\eqref{eq:nu_bar} as a consequence of the exponentially large fugacity. Conversely, T2 processes render melting possible even at low activities, where the core energy is positive and the fugacity exponentially small. This latter regime is therefore more similar, albeit only in terms of positional correlations, to the passive KTNHY melting scenario and $\bar{\nu}$ is close to its lower bound.

Before concluding, we briefly discuss one last hypothetical scenario that, while unaccessible to our two-dimensional simulations, could be possibly realized in open tissues once cell intercalation becomes negligible. In this case $c_{2}=0$ in Eq.~\eqref{eq:rg} and the RG flow occurs exclusively along the $y$-axis, with $x=x_{0}={\rm const}$. The latter dramatically affects the nature of the critical point, which is no longer in the realm of the KTNHY melting scenario. Conversely, Eq.~(\ref{eq:rg}b) is now formally identical to that governing the RG flow of a variety of spin models, including the Ising model, near the lower critical dimension (see e.g. Refs.~\cite{Brezin:1976a,Brezin:1976b,Lubensky:1978}). For $x_{0}<0$, increasing activity drives a phase transition between two solid phases, with the large activity phase -- i.e. $y_{0}>(c_{1}/c_{3})|x_{0}|$ -- characterized by the occurrence of perpetual intrusion and extrusion events, which change the nature of positional order without affecting the system mechanical properties. Under these conditions, the correlation length exhibits a more classic power-law divergence of the form: $\xi_{+}\sim|\sigma-\sigma_{c}|^{-\nu}$, with $\nu=c_{1}|x_{0}|$. This transition is instead suppressed in the liquid phase, when $x_{0}>0$ and any finite activity gives rise to an extrusion or intrusion event.

In conclusion, we have theoretically investigated the physical mechanisms facilitating the onset of collective cell migration in confluent epithelial layers. Combing cell-resolved numerical simulations and the Renormalization Group, we have shown how this fundamental process of multicellular organisms shares many similarities with defect-mediated melting in two-dimensional particle systems, but also exhibits remarkable differences, which renders the process substantially more versatile than in standard inanimate matter. Mechanical activity enters in the process by enabling states of negative defect core energy, thus rendering the solid phase unstable to cell intercalation. Confluency, on the other hand, reduces the degeneracy of dislocation unbinding, thus affecting how material parameters are renormalized across length scales. These two peculiarities, in turn, conspire towards creating three different transitional scenarios, each one characterized by a specific {\em range} of critical exponents, depending on the magnitude of fluctuations and on whether the cell layer is closed or open to intrusion and extrusion events. When cell intercalation is entirely suppressed, our approach predicts the existence of an additional phase transition between two solid phases, with the disordered phase being characterized by the occurrence of perpetual intrusion and extrusion events. The classification of different physical mechanisms via different critical exponents, $\bar{\nu}$ and $\nu$ in this case, is expected to be especially useful for a systematic and quantitative comparison with {\em in vitro} experiments.

\section{ACKNOWLEDGMENTS}
This work is supported by the ERC-CoG grant HexaTissue. D.K. acknowledges support from the NSF-DMREF-2324194 grant. Part of the computational work was carried out on the Dutch national e-infrastructure with the support of SURF through the Grant EINF-9090 for computational time, and part using the ALICE compute resources provided by Leiden University. We thank Max Bi for fruitful discussions.

\bibliography{Biblio.bib}
\end{document}


\title{Collective migration and topological phase transitions in confluent epithelia\\Supplementary information}

\author{Leonardo Puggioni}
\affiliation{Instituut-Lorentz, Universiteit Leiden, P.O. Box 9506, 2300 RA Leiden, The Netherlands}
\author{Dimitrios Krommydas}
\affiliation{Department of Physics, University of California Santa Barbara, CA 93106, USA}
\affiliation{Instituut-Lorentz, Universiteit Leiden, P.O. Box 9506, 2300 RA Leiden, The Netherlands}
\author{Luca Giomi}
\email{giomi@lorentz.leidenuniv.nl}
\affiliation{Instituut-Lorentz, Universiteit Leiden, P.O. Box 9506, 2300 RA Leiden, The Netherlands}

\maketitle

\section{Active Flow Of Cell Intercalation}

In this supplementary section we provide an analytical derivation for the active flow during Cell Intercalation. Due to the correspondence demonstrated in Ref.~[\href{https://arxiv.org/abs/2307.12956#}{21}], this active flow is the hydrodynamic flow sourced by a hexatic disclination quadrupole. 

\subsection{Scalar order parameter}

Let $\Psi_{6}=|\Psi_{6}|e^{6i\theta}$ be the hexatic complex order parameter and consider a quadrupole of two $+1/6$ and two $+1/6$ disclinations equidistantly placed from the center of the system. The phase $\theta=\theta(\bm{r})$ is then given by the convolution of the average orientation in the surroundings of each defect, that is
\begin{equation}
\label{eq:theta_convolution}
 \theta 
 = -\frac{1}{6} \arctan\left(\frac{y}{x-\ell}\right) 
 - \frac{1}{6} \arctan\left(\frac{y}{x+\ell}\right)
 + \frac{1}{6} \arctan\left(\frac{y+\ell}{x}\right) 
 + \frac{1}{6} \arctan\left(\frac{y-\ell}{x}\right)\;,
\end{equation}
where $\ell$ is distance from the center. The quadrupolar distance $\ell$ is by definition taken to be small compared to the size of the system. Thus, expanding Eq.~\eqref{eq:theta_convolution} for $|\bm{r}|/\ell \gg 1$, we obtain the simpler expression
\begin{equation}
\label{theta quad exp}
\theta = -\frac{2\ell^2\sin 2\phi}{3|\bm{r}|^2} + \mathcal{O}\left(|\ell/\bm{r}|^{6}\right)\;.
\end{equation}
Notice that the Taylor expansion features {\em only} the quadrupolar term of order $\ell^{2}$ and is exact up to $6-$th order in $\ell/|\bm{r}|$; i.e. the dipolar term, of order $\mathcal{O}(\ell/|\bm{r}|)$, and all other terms up to the $6-$th order vanish identically. 

This result is extremely robust, and can be derived in a number of ways. For instance, Eq.~\eqref{eq:theta_convolution} can be obtained from the solution of the Poisson equation
\begin{equation}\label{eq:poisson}
\nabla^{2}\varphi = \rho_{\rm d}
\end{equation}
where $\varphi$ is a dual field such that $\partial_{i}\theta=-\epsilon_{ij}\partial_{j}\varphi$ and the right-hand side of Eq.~\eqref{eq:poisson} is analogous to the electrostatic charge density [\href{https://doi.org/10.1017/CBO9780511813467}{58}]. At large distance from the defects, Eq.~\eqref{eq:poisson} can be solved by multipole expansion [\href{https://doi.org/10.1119/1.19136}{59}], that is:
\begin{equation}
\varphi = a_{0}\log \frac{r_{0}}{|\bm{r}|}+\sum_{n=1}^{\infty}\frac{a_{n}\cos n\theta+b_{n}\sin n\theta}{|\bm{r}|^{n}}\;,
\end{equation}
where $r_{0}$ is an irrelevant length scale and $a_{n}$ and $b_{n}$ are coefficients given by
\begin{subequations}\label{multipole coefficients}
\begin{gather}
a_n= \frac{1}{n} \int {\rm d}A\,|\bm{r}|^n \cos{(n\phi)} \rho_{\rm d}\;, \\[5pt]
b_n= \frac{1}{n} \int {\rm d}A\,|\bm{r}|^n \sin{(n\phi)} \rho_{\rm d}\;.
\end{gather}
\end{subequations}
Thus, up to the quadrupole term, the expansion of $\varphi$ is given by
\begin{equation}\label{diretor expansion}
\varphi = a_0 \log{\frac{r_{0}}{|\bm{r}|}} + \frac{a_1 \cos{\phi} + b_1 \sin{\phi}}{|\bm{r}|}+
\frac{a_2 \cos{2\phi} + b_2 \sin{2\phi}}{|\bm{r}|^2} + \cdots
\end{equation}
As in electrostatics, the density $\rho_{\rm d}$ is given by
\begin{equation}
\label{defect density T1}
\rho_{\rm d} = \frac{1}{6} \Big[- \delta(\bm{r} - \ell\bm{e}_{x}) - \delta(\bm{r} + \ell\bm{e}_{x}) + \delta(\bm{r} - \ell\bm{e}_{y}) + \delta(\bm{r} + \ell\bm{e}_{y}) \Big]\;,
\end{equation}
where $\ell$ is again the distance from the center of the quadrupole. Now, because the defect quadrupole has, by construction, vanishing total strength and dipole moment, $a_{0}=0$ and $a_{1}=b_{1}=0$. Of the quadrupolar terms, on the other hand, $a_{2}=-  \ell^{2}/3$ and $b_{2}=0$, thus
\begin{equation}
\label{theta quad}
\varphi = - \frac{ \ell^{2}\cos{2\phi} }{3|\bm{r}|^2}\;.
\end{equation}
Finally, going from $\varphi$ to the original field $\theta$ one finds
\begin{equation}
\label{real theta quad?}
\theta = -\frac{2 \ell^2 \sin{2\phi} }{3|\bm{r}|^2}\;,
\end{equation}
thus confirming the expression given in Eq.~\eqref{theta quad exp}.

\subsection{Active Force}

To shed light on the structure of the cellular flow triggered by a T1 process, we solve the Stokes equation in the presence of an active force of the form $\fa=\nabla\cdot\Sa$, where 
\begin{equation}\label{eq:active hexatic stress}
\Sa = \alpha_6 \nabla^{\otimes 4} \odot \boldsymbol{Q}_6,
\end{equation}
is the active hexatic stress tensor introduced in Ref.~[\href{https://doi.org/10.7554/eLife.86400}{49}]. Calculating the divergence of Eq.~\eqref{eq:active hexatic stress} yields
\begin{equation}
\label{eq:active_force}
    \fa = 960\,\frac{\alpha_{6}\ell^{2}}{|\bm{r}|^{7}}
\Bigg\{ \Bigg[
-3\cos 7\phi+\frac{\ell^2}{|\bm{r}|^{2}}\left(3\cos 5\phi-14\cos 9\phi\right)
\Bigg]\bm{e}_{x}
+\Bigg[
3\sin 7\phi-\frac{\ell^2}{|\bm{r}|^{2}}\left(3\sin 5\phi-14\sin 9\phi\right)
\Bigg]\bm{e}_{y}
\Bigg\}\;,
\end{equation}
up to correction of order $\mathcal{O}(|\ell/r|^{6})$. A plot of the force field is shown Fig.~\ref{Fig:fig_flow}a.

\subsection{Flow field}

\begin{figure}[t]
\centering 
\includegraphics[width=\textwidth]{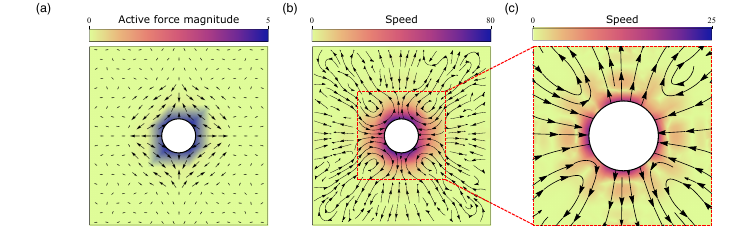}
\caption{Active stagnation flow caused by cell intercalation. (a) Force field: Stream-density plot of the force field Eq.~\eqref{eq:active_force}. It exhibits a clear, local, convergent-extension pattern in the vicinity of the quadrupolar radius $\ell$. (b) Velocity field: Stream density plot of the velocity field Eq.~\eqref{eq:velocity_full}. It exhibits a clear, local, convergent-extension flow pattern in the vicinity of the quadrupolar radius $\ell$. (c) Velocity field approximated close to defect core: Stream density plot of the velocity field Eqs.~\eqref{eq:velocity_approx}. It exhibits a clear, local, convergent-extension flow pattern in the vicinity of the quadrupolar radius $\ell$. In all plots, the light blue disk corresponds the the radius of the quadrupole. Our analytical solution is valid outside the disk.}
\label{Fig:fig_flow}
\end{figure}

To calculate the flow generated by defect quadrupole, we solve the incompressible Stokes equation sourced by the active force in Eq.~\eqref{eq:active_force}, i.e.
\begin{subequations}
\begin{gather}
\eta\nabla^{2}\bm{v}-\nabla P + \fa = \bm{0}\;,\\[5pt]
\nabla\cdot\bm{v} = 0\;,
\end{gather}	
\end{subequations}
where $\eta$ is the shear viscosity and $P$ the pressure. To this end, we turn to the Oseen formal solution
\begin{equation}
\label{Oseenv}
\bm{v}(\bm{r})= \int_{0}^{2\pi}{\rm d}\phi'\int_{\ell}^{R} {\rm d}r'r'\,\bm{G}(\bm{r}-\bm{r}') \cdot \fa(\bm{r}')\;,
\end{equation}
where
\begin{equation}
\label{OseenTensor}
\bm{G}(\bm{r}) =
\frac{1}{4\pi \eta} \Bigg[\left(\log{\frac{\mathcal{L}}{|\bm{r}|}}-1\right)\mathbb{1} +
\frac{\bm{r}\otimes\bm{r}}{|\bm{r}|^2} \Bigg]\;,
\end{equation}
is the two-dimensional Oseen tensor, with $\mathcal{L}$ a constant, and $R$ is a large distance cut-off. Without loss of generality, one can set $\mathcal{L}= R\sqrt{e}$ in Eq.~\eqref{OseenTensor}. To calculate the integrals in Eq.~\eqref{Oseenv}, we make use of the logarithmic expansion
\begin{equation}
\label{logexp}
     \log{\frac{|\bm{r}-\bm{r}'|}{\mathcal{L}}} = \log{\frac{r_>}{\mathcal{L}}} - \sum^\infty_1 \frac{1}{m}\left(\frac{r_>}{r_<}\right)^m \cos{[m(\phi-\phi')]}\;,
\end{equation}
with $r_{\gtrless}$ the maximum (minimum) between $|\bm{r}|$ and $|\bm{r}'|$, and of the orthogonality of trigonometric functions 
\begin{equation}
\int^{2\pi}_0 \text{d}\phi'  \cos{[m(\phi-\phi')]}\cos{n\phi'} = \pi \cos{n\phi}~ \delta_{mn}\;.
\end{equation}
The resulting flow field surrounding the defect quadrupole is then given by 
\begin{align}
\frac{\bm{v}}{2\alpha_6 \ell^2/\eta}  
&=-\Bigg[\frac{30\left(\ell^2-\bm{|r|}^2\right)^2}{\bm{|r|}^{12}} \left(3 \bm{|r|}^2 \cos {8\phi} + 14 \ell^2 \cos{10\phi}\right)
+\frac{60 }{\bm{|r|}^{8}}\cos {6\phi} \left(3\bm{|r|}^2+6 \ell^2 \log \frac{\bm{|r|}}{\ell}-4 \ell^2\right)\Bigg]\bm{r}\notag\\[5pt]
&+6\left(\frac{6 }{\bm{|r|}^5}-\frac{5 \ell^2}{\bm{|r|}^{7}}\right) (\cos{5\phi}~\bm{e}_x-\sin{5\phi}~\bm{e}_y) 
+\frac{30}{7 } \left(\frac{6 \ell^2}{\bm{|r|}^7}-\frac{7}{\bm{|r|}^5}\right) (\cos{7\phi}~\bm{e}_x-\sin{7\phi}~\bm{e}_y)\notag\\[5pt]
&+\frac{35}{3} \ell^2 \left(\frac{8 \ell^2}{\bm{|r|}^9}-\frac{9}{\bm{|r|}^7}\right)  (\cos{9\phi}~\bm{e}_x-\sin{9\phi}~\bm{e}_y)\;.
\label{eq:velocity_full}
\end{align}
The flow field in Eq.~\eqref{eq:velocity_full} quite complicated. The magnitude of the flow and the active quadrupolar force that sources it, however, is, as it should, much larger in the vicinity of the defect quadrupole (Fig.~\ref{Fig:fig_flow}a,b). Indeed, in close proximity of the defect quadrupole Eq.~\eqref{eq:velocity_full} becomes
\begin{equation}\label{eq:velocity_approx}
\bm{v} \approx \frac{120\alpha_{6}\ell^{4}}{\eta} \left[\left(4-6\log{\frac{\bm{|r|}}{\ell}}-3\,\frac{\bm{|r|}^2}{\ell^{2}}\right) \cos6\phi \right] \frac{\bm{r}}{|\bm{r}|^{8}}\;.
\end{equation}
In this region, i.e. $|\bm{r}|/\ell \gtrapprox 1$, the flow Eq.~\eqref{eq:velocity_approx} has the typical stricture of a stagnation flow (Fig.~\ref{Fig:fig_flow}c). That is, the components of the velocity field can be written as $v_{x}(x,0) \sim \alpha_{6}/(\eta \ell^{4})\,x$ and $v_{y}(y,0) \sim - \alpha_{6}/(\eta\ell^{4})\,y$. This result can be used to construct the viscous stress in Eq.~(1) of the main text.

\section{Analysis of correlation}

In this supplementary section we provide a detailed derivation of the analytical results in the main text, concerning the behavior of spatial correlations of a confluent monolayer in the {\em active} phase.  Following the classic approach pioneered by Kosterlitz and Thouless~~[\href{https://doi.org/10.1088%2F0022-3719%2F5%2F11%2F002}{14}, \href{https://doi.org/10.1088/0022-3719/7/6/005}{15}], Halperin and Nelson~[\href{https://doi.org/10.1103/PhysRevB.19.2457}{16}] and Young~[\href{https://doi.org/10.1103%2FPhysRevB.19.1855}{17}], this is achieved by means of a Renormalization Group (RG) analysis of a gas of dislocations, whose interactions is governed by the Hamiltonian
\begin{equation}
\mathcal{H}_{D} = -\frac{Y}{8\pi k_{B}T}\sum_{i \ne j} \bm{G}(\bm{r}_{i}-\bm{r}_{j}):(\bm{b}_{i}\otimes\bm{b}_{j})+\frac{\epsilon_{c}+\sigma}{k_{B}T}\sum_{i}|\bm{b}_{i}|^{2}\;,
\end{equation}
where $Y$ is the Young modulus of the crystal, $\bm{r}_{i}$ the position of the $i$-th dislocation with Burgers vector $\bm{b}_{i}$ and, as explained in the main text, $\epsilon_{c}+\sigma$ is the effective core energy per unit area of dislocations. The tensor
\begin{equation}
\bm{G}(\bm{r}) = \log\left(\frac{|\bm{r}|}{a}\right)\mathbb{1}-\frac{\bm{r}\otimes\bm{r}}{|\bm{r}|^{2}}\;, 	
\end{equation}
is the elastic Green function mediating the interactions between dislocations, with the colon indicating the Euclidean product of two tensors (i.e. $\bm{A}:\bm{B}=A_{\alpha\beta}B_{\alpha\beta}$). Following Refs~[\href{https://doi.org/10.1103/PhysRevB.19.2457}{16},\href{https://doi.org/10.1103%2FPhysRevB.19.1855}{17}], this program yields the following recursion relations for the variables $x=K^{-1}-K_{c}^{-1}$, with $K=Ya^{2}/(k_{B}T)$ and $K_{c}=16\pi$ the critical stiffness, and the fugacity $y=e^{-(\epsilon_{c}+\sigma)a^{2}/(k_{B}T)}$. That is
\begin{subequations}\label{eq:rg}
	\begin{gather}
	\frac{{\rm d}x}{{\rm d}l} = c_{2}y^{2}\;,\\
	\frac{{\rm d}y}{{\rm d}l} = c_{1}xy+c_{3}y^{2}\;, 	
	\end{gather}
\end{subequations}
where ${\rm d}l=\log b$, with $b$ the RG scaling factor, and $c_{1}$, $c_{2}$ and $c_{3}$ positive constants, reflecting different aspects of dislocation unbinding. Specifically the constant $c_{1}$ results from the competition between the energetic cost and the entropic gain associated with the creation of a single dislocation and, in a two-dimensional crystal of point particles is given by $c_{1}=2K_{c}=32\pi$. The constants $c_{2}$ and $c_{3}$, on the other hand, embody two different types of unbinding processes, involving the simultaneous creation of two and three dislocations with vanishing net Burgers vector: i.e. $\sum_{i}\bm{b}_{i}=\bm{0}$. In a six-fold coordinated crystal these are given by $c_{2}=12\pi^{2}e^{2}[2I_{0}(2)-I_{1}(2)]=2597.84\ldots$ and $c_{3}=2\pi e I_{0}(2)=38.93\ldots$, with $I_{0}$ and $I_{1}$ modified Bessel functions~[\href{https://doi.org/10.1103/PhysRevB.19.2457}{16},\href{https://doi.org/10.1103%2FPhysRevB.19.1855}{17}]. 
			
Now, while moving from point particles in a crystal to cells in a confluent monolayer, the specific magnitude of $c_{1}$, $c_{2}$ and $c_{3}$ is expected to shift away from the classic estimate by Halperin, Nelson and Young. Yet, as long as the transition occurs through the unbinding of dislocations, the RG flow preserves the fundamental structure expressed by Eqs.~\eqref{eq:rg}. Taking advantage of this, in the following sub-sections we will detail the three scenario discussed in the main text, where the transition is driven solely by T1 processes (Sec.~\ref{sec:pure_t1}), by both T1 and T2 processes (Sec.~\ref{sec:t1_and_t2}) and by T2 processes only (Sec.~\ref{sec:pure_t2}), while treating $c_{1}$, $c_{2}$ and $c_{3}$ as material parameters.
			
\subsection{\label{sec:pure_t1}Case I: pure T1}
			
noindent When T1 processes are the sole topological excitations in a cell monolayer, $c_{3}=0$ and Eqs.~\eqref{eq:rg}
\begin{subequations}\label{eq:rg_kt}
	\begin{gather}
	\frac{{\rm d}x}{{\rm d}l} = c_{2}y^{2}\;,\\
	\frac{{\rm d}y}{{\rm d}l} = c_{1}xy\;,
	\end{gather}
\end{subequations}
from which one can now readily show that 
\begin{equation}
\frac{{\rm d}}{{\rm d}l}\left(x^{2}-\frac{c_{2}}{c_{1}}\,y^{2}\right) = 0\;.	
\end{equation}
The latter implies that the quantity $\mathcal{C}=x^{2}-(c_{2}/c_{1}) y^{2}$ is conserved along the RG flow. As a consequence, the flow occurs along hyperbolas, whose foci are located on either the $x-$ or $y-$axis depending on the sign of $\mathcal{C}$. For $\mathcal{C}>0$, the foci lie on the $x-$axis and the trajectories flow toward the solid phase: i.e. $y\to 0$. For $\mathcal{C}<0$, on the other hand, the foci are located on the $y-$axis and the RG flow converges towards the liquid phase: i.e. $(x,y)\to(\infty,\infty)$. Finally, the condition $\mathcal{C}=0$ defines the asymptotes, i.e. $x=\pm \sqrt{c_{2}/c_{1}}\,y$, separating these two families of trajectories on the $xy$-plane. 
			
To compute the critical exponent $\bar{\nu}$, we assume approaching the transition from the liquid phase, so that $\mathcal{C}<0$. Then, taking $y^{2}=(c_{1}/c_{2})(x^{2}+|\mathcal{C}|)$ and replacing it in Eq.~(\ref{eq:rg_kt}a) yields a single differential equation of the form
\begin{equation}
\frac{{\rm d}x}{{\rm d}l} = c_{1}(x^{2}+|\mathcal{C}|)\;,
\end{equation}
whose solution in the range $l\ge 0$ can now be readily expressed in the form
\begin{equation}\label{eq:kt_sol}
\frac{\arctan\frac{x(l)}{\sqrt{|\mathcal{C}|}}}{\sqrt{|\mathcal{C}|}}-\frac{\arctan\frac{x_{0}}{\sqrt{|\mathcal{C}|}}}{\sqrt{|\mathcal{C}|}} = c_{1}l\;,
\end{equation}
where $x_{0}=x(0)$. Eq.~\eqref{eq:kt_sol} can now be used to estimate the length scale at which fluctuations no longer drives a significant renormalization of the material parameters. The latter, in turn, sets the the magnitude of the correlation length $\xi_{+}$. That is 
\begin{equation}
\xi_{+} \approx a e^{l^{*}}\;,	
\end{equation}
where $l^{*}$ marks the point along the RG flow where $\max\{x^{*}y^{*},y^{*\,2}\}=1$, with $x^{*}=x(l^{*})$ and $y^{*}=y(l^{*})$. Where either $xy>1$ or $y^{2}>1$, thus for $l>l^{*}$, the truncation leading to Eqs.~\eqref{eq:rg_kt} is no longer valid. This implies $x^{*}\approx y^{*} \approx 1$, thus $\arctan(x^{*}/\sqrt{|\mathcal{C}|}) \approx \pi/2$ in proximity of the critical point, where $|\mathcal{C}|\approx 0$. From these considerations and Eq.~\eqref{eq:kt_sol} it follows that
\begin{equation}\label{eq:ell_star}
l^{*} 
\sim
\left\{
\begin{array}{lll}
|\mathcal{C}|^{-1/2} 	& & |\mathcal{C}| \ll |x_{0}|\;, \\[5pt]
|\mathcal{C}|^{-1}		& & |\mathcal{C}| \gg |x_{0}|\;.	
\end{array}
\right.
\end{equation}
Now, in passive hexatics, where $\sigma=0$ and temperature is the sole parameter driving the transition, $\mathcal{C} \sim (T_{c}-T)$ up to a positive constant. Furthermore, assuming $\epsilon_{c}a^{2}/k_{B}T \approx 1$ implies $|x_{0}| \gg |\mathcal{C}| \ge 0$ near criticality, thus
\begin{equation}\label{eq:xi_eq}
\log\left(\xi_{+}/a\right) \sim |T-T_{c}|^{-1/2}\;,
\end{equation}
in all practical circumstances. In active cell layers, on the other hand, the same hierarchy of energy scales is no longer guaranteed in case of extensile activity (i.e. $\sigma<0$). Taking $\mathcal{C} \sim \sigma_{c}-\sigma$, with $\sigma_{c}$ a critical active stress, Eq.~\eqref{eq:ell_star} now implies
\begin{equation}
\log\left(\xi_{+}/a\right) \sim |\sigma-\sigma_{c}|^{-\bar{\nu}}\;,
\end{equation}
where $\bar{\nu}$ is a non-universal exponent in then range
\begin{equation}\label{eq:nu_bar_range}
\frac{1}{2} \le \bar{\nu} \le 1\;.
\end{equation}
The derivation summarized here does not provide a clear indication about where, in the range given by Eq.~\eqref{eq:nu_bar_range}, is more likely to find $\bar{\nu}$. The latter will become evident in the next section.
			
\subsection{\label{sec:t1_and_t2}Case II: T1 and T2}
			
When T2 processes are also at play, $c_{3}>0$ and Eqs.~\eqref{eq:rg_kt} have the same structure investigated in Refs.~[\href{https://doi.org/10.1103/PhysRevB.19.2457}{16}, \href{https://doi.org/10.1103%2FPhysRevB.19.1855}{17}] for two-dimensional melting. In spite of the lack of an integral of motion, the structure of the RG flow is analogous to that discussed in the previous sub-section, with a separatrix of negative slope partitioning the $xy$-plane into a solid and a liquid phase. To calculate this slope one can set $y=mx$ and reduce Eqs.~\eqref{eq:rg} to the single quadratic equation
\begin{equation}
c_{2}m^{2}-c_{3}m-c_{1} = 0\;,	
\end{equation}
whose solutions are readily found in the form
\begin{equation}
m_{\pm} = \frac{c_{3}\pm \sqrt{c_{3}^{2}+4c_{1}c_{2}}}{2c_{2}}\;.
\end{equation}
To investigate the divergence of the correlation length as the critical point is approached from the disordered phase, one can consider trajectories whose origin $(x_{0},y_{0})$ is sufficiently close to the separatrix, that is $y=m_{-}x+D$, with $D=D(l)$ a deviation along the $y-$axis. For $D=0$, replacing this in Eqs.~\eqref{eq:rg} yields the RG flow along the separatrix, thus for $\sigma=\sigma_{c}$: i.e.
\begin{subequations}\label{eq:separatrix}
	\begin{gather}
	x(l) = \frac{x_{0}}{1-c_{2}x_{0}m_{-}^{2}l}\;,\\[5pt]
	y(l) = \frac{m_{-}x_{0}}{1-c_{2}x_{0}m_{-}^{2}l}\;.
	\end{gather}
\end{subequations}
Replacing this in Eqs.~\eqref{eq:rg} gives after standard manipulations
\begin{equation}\label{eq:dddl}
\frac{{\rm d}D}{{\rm d}l} = -c_{1} x D+ c_{2}m_{+}D^{2}\;.
\end{equation}
Now, in the classic treatment of the melting transition, the second term on the right-hand side of this equation is neglected under the assumption that $|c_{1}x| \gg |m_{+}D|$ in the range $0\le l \le l^{*}$. This yields
\begin{equation}\label{eq:d_sol}
D(l)
\approx D_{0}\left(1+c_{2}|x_{0}|m_{-}^{2}l\right)^{\frac{c_{1}}{c_{2}m_{-}^{2}}}\;
=D_{0}\left[\frac{x_{0}}{x(l)}\right]^{\frac{c_{1}}{c_{2}m_{-}^{2}}}\;,
\end{equation}
where $D_{0}=D(0)$ and, as in Sec.~\ref{sec:pure_t1} we made explicit that $x_{0}<0$. This equation can now be used, in alternative to Eq.~\eqref{eq:kt_sol}, to compute $l^{*}$, hence the correlation length. Specifically, Eq.~\eqref{eq:dddl} is no longer valid when $|D^{*}| \approx |m_{-}x^{*}|$, with $D^{*}=D(l^{*})$, from which, using Eq.~\eqref{eq:d_sol}, one readily finds
\begin{equation}
x^{*} \sim D_{0}^{\frac{c_{2}m_{-}^{2}}{c_{2}m_{-}^{2}+c_{1}}}\;.
\end{equation}
Finally, taking into account that $x^{*} \sim 1/l^{*} \sim |\sigma-\sigma_{c}|^{\bar{\nu}}$ and $D_{0} \sim (\sigma-\sigma_{c})$ we conclude that the exponent $\bar{\nu}$ is now given by 
\begin{equation}\label{eq:nu_t2}
\bar{\nu} 
= \frac{c_{2}m_{-}^{2}}{c_{2}m_{-}^{2}+c_{1}}
= \frac{1}{2}\left(1-\frac{c_{3}}{\sqrt{c_{3}^{2}+4c_{1}c_{2}}}\right)\;.
\end{equation}
As in Sec.~\ref{sec:pure_t1}, driving the transition by means of the cell speed $v$ rather than temperature necessitates extending this analysis to the case where both terms on the right-hand side of Eq.~\eqref{eq:dddl}. To avoid unnecessary complications, this can be achieved by considering the opposite limit, in which $|c_{1}x|\ll|m_{+}D|$ and solving Eq.~\eqref{eq:dddl} gives
\begin{equation}
\frac{1}{D_{0}}-\frac{1}{D(l)} \approx c_{2}m_{+}l\;. 
\end{equation}
By construction, $m_{+}>0$ and $|D(l^{*})|\gg |D_{0}|$, so that $l^{*} \sim D_{0}^{-1}$ and 
\begin{equation}
\bar{\nu} = 1\;.	 
\end{equation}
From this we conclude that the effect of cell extrusion/intrusion on the transition to collective migration is to further extend the range of the $\bar{\nu}$ exponent, so that
\begin{equation}\label{eq:nu_bar_range}
\frac{1}{2}\left(1-\frac{c_{3}}{\sqrt{c_{3}^{2}+4c_{1}c_{2}}}\right) \le \bar{\nu} \le 1\;.
\end{equation}
The specific value of $\bar{\nu}$ in the range given by Eq.~\eqref{eq:nu_bar_range} depends, therefore, on the distance along the $y$-axis between the trajectory emanating from the initial condition $(x_{0},y_{0})$ and the separatrices, here expressed by the function $D$. In passive systems, where temperature is the sole parameter driving the melting transition and the core energy is positive, the initial fugacity is exponentially small, hence $0 \le y_{0} \ll 1$. This implies that $(x_{0},y_{0})$ lies in proximity of the origin of the $xy$-plane near criticality and that the trajectory originating from it remains close to the separatrix of the $x>0$ quadrant. In {\em closed} epithelial layers, on the other hand, melting is generally driven upon increasing the activity from the solid phase, thus $x_{0}<0$ and $y_{0} \approx 1$. The trajectory emanating from this initial condition is thus generally further away the separatrix than in passive systems and $\bar{\nu}$ is closer to the upper bound of its range. In {\em open} epithelial layers, by contrast, T2 events soften the system even when modestly active and the initial fugacity is again exponentially small. The exponent $\bar{\nu}$ resulting from this process is, therefore, expected to be close to the lower bound of Eq.~\eqref{eq:nu_bar_range}.
		 	
\subsection{\label{sec:pure_t2}Case III: pure T2}
				
To complete this analysis, we consider the case where the transition is driven solely by T2 processes. In this case $c_{2}=0$ and $x=x_{0}={\rm const}$ along the RG flow. By contrast, Eq.~(\ref{eq:rg}b) reduces to the simple autonomous equation
\begin{equation}\label{eq:rg_t2}
\frac{{\rm d}y}{{\rm d}l} = c_{1}x_{0}y+c_{3}y^{2}\;.
\end{equation}
Remarkably, the latter is formally identical to the equation describing the renormalization of temperature in a variety of classic models of statistical physics in $d_{l}+\epsilon$ dimensions, with $d_{l}$ the lower critical dimensions. In the Ising model, for instance, $d_{l}=1$ and the temperature $T$ renormalizes according to equation
\begin{equation}\label{eq:ising}
\frac{{\rm d}T}{{\rm d}l} =-\epsilon T + \frac{1}{2}\,T^{2}\;.
\end{equation} 
The same equation holds for the non-linear sigma model~[\href{https://doi.org/10.1103/PhysRevLett.36.691}{55},\href{https://doi.org/10.1103/PhysRevB.14.3110}{56}] and the zero state Potts model in $2+\epsilon$ dimensions~[\href{https://doi.org/10.1016/0375-9601(78)90479-6}{57}], with $d_{l}=2$ in both cases. In this analogy, $\epsilon=d-d_{l}$ determines whether a system can transition to an ordered phase at finite temperature. Specifically, for $\epsilon<0$, hence below the lower critical dimension, Eq.~\eqref{eq:ising} has no critical fixed point in the physical range $T>0$, whereas for $\epsilon>0$, the system exhibits long-ranged order for $T<T_{c}=2\epsilon$.
				
In the case of confluent cell layers, $\epsilon \leftrightarrow -c_{1}x_{0}$ and the same behavior is recovered depending on the sign of the stiffness parameter $x_{0}$. If the monolayer is initially in a solid phase, where $x_{0}<0$, Eq.~\eqref{eq:rg_t2} admits a critical fixed point when $y_{c}=c_{1}|x_{0}|$. Unlike in the previous two cases, density correlation here exhibit a more standard power-law behavior in the vicinity of the critical point: i.e.
\begin{equation}
\xi \sim |\sigma-\sigma_{c}|^{-\nu}\;,
\end{equation}  
with $\alpha_{c}$ a critical activity associated with $y_{c}$ and $\nu$ another non-universal exponent. This can be easily determined by linearizing Eq.~\eqref{eq:rg_t2} near the point, to give 
\begin{equation}
\nu=c_{1}|x_{0}|\;.
\end{equation}
Conversely, when $x_{0}>0$, the transition disappears and any finite activity gives rise to an extrusion or intrusion event.

\section{Multiphase-field model}

The numerical results shown in the main text are obtained by integrating of a two-dimensional version of the model introduced in Refs.~[\href{https://doi.org/10.7554/eLife.82435}{43}, \href{https://doi.org/10.1098/rsif.2024.0022}{44}]. Each cell $i$ is described by a continuous phase-field $\phi_{i}=\phi_i \left({ \bm r}, t \right)$, obeying to the equation:
\begin{equation}\label{eq:mpf}
\partial_t \phi_i + \bm{v}_i \cdot \nabla \phi_i = - \frac{\delta \mathcal{F}}{\delta \phi_i}, \qquad \qquad i=1,\,2 \dots\,N\;,
\end{equation}
where $\bm{v}_i$ is the velocity of the cell and $\mathcal{F}=\mathcal{F}\left[\phi\right]$ a free-energy functional, describing the passive behavior of the $N$ cells:
\begin{align}\label{eq:free}
\mathcal{F} \left[ \{\phi_i \} \right] = & \sum_i^N \frac{E}{\lambda} \int {\rm d}^{2}r\, \left[ 4 \phi_i^2 \left( 1 - \phi_i \right)^2 + \lambda^2 \left( \nabla \phi_i \right)^2 \right] + \sum_i^N \mu \left( 1 - \frac{1}{A_0} \int {\rm d}^{2}r\, \phi_i^2 \right)^2 + \nonumber \\ &+ \sum_i^N \sum_{j \neq i} \frac{\kappa}{\lambda} \int {\rm d}^{2}r\, \phi_i^2 \phi_j^2 + \sum_i^N \sum_{j \neq i} \frac{\omega}{\lambda^2} \int {\rm d}^{2}r\, \nabla \phi_i \cdot \nabla \phi_j\;.
\end{align}
The first term in \eqref{eq:free} is a Cahn-Hilliard potential minimizing the cell perimeter, the second one is a soft constraint on the area around $A_0 = \pi R_0^2$, while the third and fourth ones represent, respectively, cell-cell repulsion and cell-cell boundary adhesion.
Velocity is obtained by assuming overdamped dynamics:
\begin{equation}
\bm{v}_i = v_{0} \bm{p}_i + \zeta^{-1}\int {\rm d}^{2}r\, \phi_j \nabla \left( -\sum_j^N \frac{\delta \mathcal{F}}{\delta \phi_j} \right), 
\end{equation}
where $\zeta$ is the substrate friction, $v_{0}$ the cells speed and $\bm{p}_i = \left( \cos \Theta_i, \sin \Theta_i\right)$ a unit vector expressing the direction of self-propulsion. The latter, in turn, is subject to rotational Brownian motion, so that
\begin{equation}
\frac{{\rm d} \Theta_i}{{\rm d} t} = \sqrt{ 2 D_r}\,\eta\;,
\end{equation} 
with $D_r$ the rotational diffusivity and $\eta$ a Gaussian white noise with zero mean and unit variance.

\subsection{Active stress}

Let $\bm{f}_{i}=v_{0}\bm{p}_{i}$ be the active force acting on the center of mass of the $i-$th cell, whose position is denoted by $\bm{r}_{i}$, the local coarse-grained stress field at position $\bm{r}$ can be computed, following e.g. Ref.~[\href{https://doi.org/10.1115/1.3157619}{58},\href{https://doi.org/10.1038/s42003-022-03288-x}{59}]), as
\begin{equation}
\bm{\sigma}(\bm{r}) = \frac{1}{2A_{\Omega}}\,\sum_{i\in\Omega} \left[\frac{\bm{f}_{i} \otimes (\bm{r}-\bm{r}_{i})}{|\bm{r}-\bm{r}_{i}|}  +  \frac{(\bm{r}-\bm{r}_{i})\otimes\bm{f}_{i}}{|\bm{r}-\bm{r}_{i}|} \right]\;,
\end{equation}
where $\Omega$ is a domain of area $A_{\Omega}$ centered at $\bm{r}$. From this definition, it follows that $\sigma \sim v_{0}$.

\subsection{Cell intrusion/extrusion}

In order to mimic the process of cell extrusion in a 2D configuration, the coefficient $\mu$ associated with the area constraint is not the same for all the cells, nor is speed $v_{0}$. At regular time intervals a cell is randomly selected to be extruded: the corresponding value of $\mu$ is progressively decreased from the ``living cell" value $\mu = \mu_0$ to the ``dead cell" value $\mu = \mu_1 \ll \mu_0$. The latter, renders the selected cell weaker and less able to withstand the lateral pressure exerted by neighboring cells, thus causing the area to gradually decrease until its complete disappearance. The extrusion is therefore not directly caused by the cell itself, since the target area value $A_0$ is not modified, but by the interaction with its neighbors. For this reason, the time required to completely extrude a cell depends on the particular configuration of the neighborhood at that moment, and therefore the effective extrusion/intrusion rate is not constant or known \textit{a priori}. To be consistent with the fact that extrusion is associated to apoptosis, we also set $v_{0}=0$ in the extruded cell.

In order to keep the total number of cells fixed, after the complete extrusion of a cell a new one must be injected by an intrusion process. We randomly choose a vertex $\bm{r}_v$ of the tissue (\textit{i.e.} a point where 3 or 4 neighboring cells converge), and we assign a value different from 0 to the field $\phi_i$ associated with the new cell in the points surrounding $\bm{r}_v$. The relaxation term $- \delta \mathcal{F} / \delta \phi_i$ will immediately make the new cell to grow and to push the neighbors away until it reaches a state comparable to the other ones, although sometimes the pressure from the neighbors is too large to overcome, and the process must be repeated in a different point. To obtain a smoother process, the value of $\mu$ is not immediately set to $\mu_0$, but gradually increased.

\subsection{Numerical implementation}

Eq.~\eqref{eq:mpf} is solved in a $N_x \times N_y = 496\times490$ periodic domain with $N=1085$ cells, using a GPU-accelerated finite-difference code with an Euler-Maruyama time-evolving scheme. Grid-spacing and timestep are respectively $\Delta x = \Delta y = 1$ and $\Delta t = 0.02$. The parameters we used are $E = 0.015$, $\lambda = 3$, $\mu = \mu_0 = 45$, $\mu_1=5$, $\kappa = 0.5$, $\omega = 0.0015$, $\zeta = 1$, $D_r = 0.0001$. Regarding the cell size, we considered a radius $R_0 = 12$ such that the total target area $NA_0$ is larger than the available one, in order to achieve confluency (similarly to Ref.~[\href{https://journals.aps.org/prl/abstract/10.1103/PhysRevLett.125.038003}{35}]).

Simulations are initialized by placing the cells in a triangular lattice and allowing the system to relax without activity for $10^4$ time steps. Subsequently we let the system to evolve with cell self-propulsion (and eventually with cell intrusion/extrusion) until its statistical properties become stationary, which can take up to $4\times10^6$ temporal iterations, afterwards we let the system to run for additional $2 \times 10^{6}$ time steps, in order to perform statistical averaging over time. For each value of activity we run three independent simulations, with a different initial configuration of polarization vectors $\{ \bm{p}_i\}$, a different realization of the Gaussian noise $\eta$ and, for simulations with T2 events, a different realization of the random processes selecting the position of cell intrusions and extrusions. All the results presented in the text are obtained averaging both over time in the stationary regime and over the three different realizations.

\subsubsection{Positional order}

Positional order is quantified by means of the structure factor, $S(\bm{q})$, defined from the Fourier transform of the correlation function of the cell density
\begin{equation}
\rho(\bm{r}) = \sum_{i=1}^{N}\delta(\bm{r}-\bm{r}_{i})\;.
\end{equation}
Using standard manipulations, one can readily show that
\begin{align}\label{eq:structure_factor}
S(\bm{q}) 
&= \frac{1}{N}\,\sum_{j=1}^{N}\sum_{k=1}^{N}e^{i\bm{q}\cdot(\bm{r}_{j}-\bm{r}_{k})} \notag \\
&= \frac{1}{N}\,
\int {\rm d}^{2}r\,\sum_{j=1}^{N}e^{i\bm{q}\cdot\bm{r}}\delta(\bm{r}-\bm{r}_{j})
\int {\rm d}^{2}r\,\sum_{k=1}^{N}e^{-i\bm{q}\cdot\bm{r}}\delta(\bm{r}-\bm{r}_{k}) \notag \\
&=\frac{\langle \rho_{\bm{q}}\rho_{-\bm{q}} \rangle}{N}\;,
\end{align}
where $\rho_{\bm{q}}=\int{\rm d}^2r\,e^{i\bm{q}\cdot\bm{r}}\rho(\bm{r})$. To compute the coordinates the center of mass of a given cell, we use the Bai-Breen algorithm [\href{https://doi.org/10.1080/2151237X.2008.10129266}{62}], which takes into account the periodic boundary conditions mapping each coordinate onto a circle. For instance, to extract the $x$-coordinate of the center of mass of the $i$-th cell, we define a new angular coordinate $\vartheta_n = 2\pi x_n/x_{\max}$, with $n$ running over all the gridpoints in the domain and $x_{\max} = \Delta x N_x$. We then compute the average quantities
\begin{subequations}
\begin{align}
\cos \vartheta_i 
= \frac{\sum_n \phi_i \left( \bm{r}_{n} \right) \cos \vartheta_n}{\sum_n \phi_i \left(\bm{r}_{n}\right)}\;,\\[10pt]
\sin \vartheta_i 
= \frac{\sum_n \phi_i \left( \bm{r}_{n} \right) \sin \vartheta_n}{\sum_n \phi_i \left(\bm{r}_{n}\right)}\;,
\end{align}
\end{subequations}
from which we obtain $\vartheta_i = \arctan ( \sin \vartheta_i / \cos \vartheta_{i}  ) + \pi$, and finally $x_i = \vartheta_i x_{\max} /2\pi$. The same procedure can be used to extract the $y$-coordinate. 

\subsubsection{Orientational order}

In order to investigate the orientational order and the presence of defects, we tessellate the domain assigning each point to the cell with the large value of the phase field at that point:
\begin{equation}
\bm{r}_a \in i-\text{cell} \qquad  \text{if} \quad  \phi_i \left( \bm{r}_a \right) = \max\left( \{ \phi_j \left( \bm{r}_a \right) \} \right) \qquad j=1,\,2 \dots\, N\;.
\end{equation} 
Such a tessellation is correct if the tissue confluence is satisfied, and vertices are therefore defined as the points where 3 or 4 cells converge. This further allows us to locate and count the neighbors of each cell. Thus, letting $V$ be the total number of vertices of a cell, free dislocations are defined as pairs of adjacent cells having respectively $V=5$ and $V=7$, completely surrounded by cells with $V=6$, and free disclinations as isolated cells having $V=5$ or $V=7$, similarly to previous studies employing different models of epithelial tissues~[\href{https://doi.org/10.1103/PhysRevMaterials.2.045602}{8}, \href{https://doi.org/10.1103/PhysRevLett.123.188001}{9},
\href{https://doi.org/10.1039/D0SM00109K}{10}].

Orientational order can then quantified in two different (and here equivalent) ways. From the coordinates of the center of center of mass of the neighbors we compute the local $6$-fold orientation of the virtual bond connecting the center of mass of the $j$-th cell with that of its $k$-th neighbor: i.e. $e^{6i\vartheta_{jk}}$, with $\vartheta_{jk}= \arctan[(y_j - y_k)/(x_j - x_k)]$. Then, averaging over all the neighbors, we compute the hexatic order parameter $\Psi_{6}$ at the cellular scale. That is
\begin{equation}
\Psi_{6}(\bm{r}_{j}) = \frac{1}{V_{j}}\,\sum_{k=1}^{V_{j}} e^{6i\vartheta_{jk}}\;,
\end{equation}
where $V_{j}$ is the number of vertices of the $j-$th cell, equal to the number of neighbors. The orientational correlation function is then defined as
\begin{equation}
g_6 (\bm{r}) = \langle \Psi_{6}(\bm{r})\Psi_{6}^{*}(\bm{0})\rangle\;.
\end{equation}

A different way to compute orientational correlations, which does not require introducing bonds, but leverage instead on the sole orientation of cells, was introduced in Ref.~ [\href{https://doi.org/10.1038/s41567-023-02179-0}{48}] and relies of the function
\begin{equation} 
\gamma_{6}(\bm{r}_{j}) = \frac{\sum_{v=1}^{V_{j}}|\bm{r}_{j}-\bm{r}_{v}|^{6}e^{6i\vartheta_{jv}}}{\sum_{v}|\bm{r}_{j}-\bm{r}_{v}|^{6}}\;,
\end{equation}
where the summation now runs over all the vertices of the $j$-th cell, whose position is denoted by $\bm{r}_{v}$, with $\vartheta_{jv}= \arctan[(y_j - y_v)/(x_j - x_v)]$. Next, using the polar decomposition, i.e. $\gamma_6 = |\gamma_6| e^{i 6 \theta}$, one can define two alternatives orientational correlation functions, that is
\begin{equation}
g_6^S (\bm{r}) = \langle \gamma_6 \left(\bm{r}\right) \gamma_6^{*} \left(\bm{0}\right) \rangle\;, 
\end{equation}
and the angular correlator
\begin{equation}
g_6^{\theta} \left(\bm{r}\right) = \langle e^{6i \left[\theta \left(\bm{r}\right) - \theta \left(\bm{0}\right)\right]}\rangle\;.
\end{equation}
For sufficiently monodisperse cellular layers, all these metrics convey the same picture, as one can readily verify comparing Fig.~2c of the main text with Fig.~\ref{fig:S3}.

\newpage

\section{Supplementary figures}

\begin{figure}[h!]
	\centering
	\includegraphics[width=\textwidth]{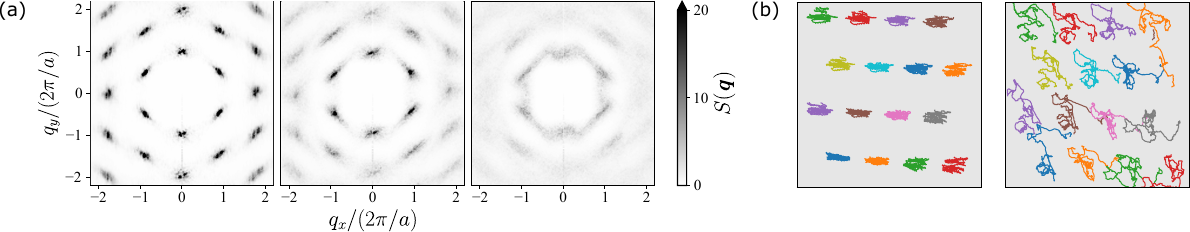}	
	\caption{Multiphase field simulations in open systems (a) Structure factor in the crystalline, hexatic and isotropic liquid phase. (b) Trajectories of a selected number of cells across the melting transition.}
\end{figure}

\begin{figure}[h!]
	\centering
	\includegraphics[width=\textwidth]{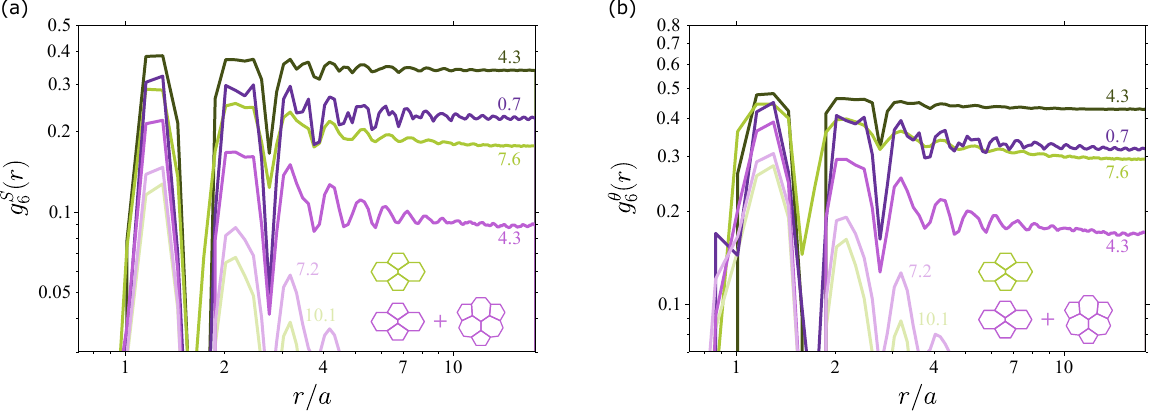}	
   \caption{Alternative definitions of the orientational correlation function in the multiphase field simulations (a) shape correlation functions, (b) angular correlation functions, in closed (green tones) and open (magenta tones) systems, for varying self propulsion speed $v_0/(aD_r)$.}
   \label{fig:S3}
\end{figure}

\begin{figure}[h!]
	\centering
	\includegraphics[width=0.45\textwidth]{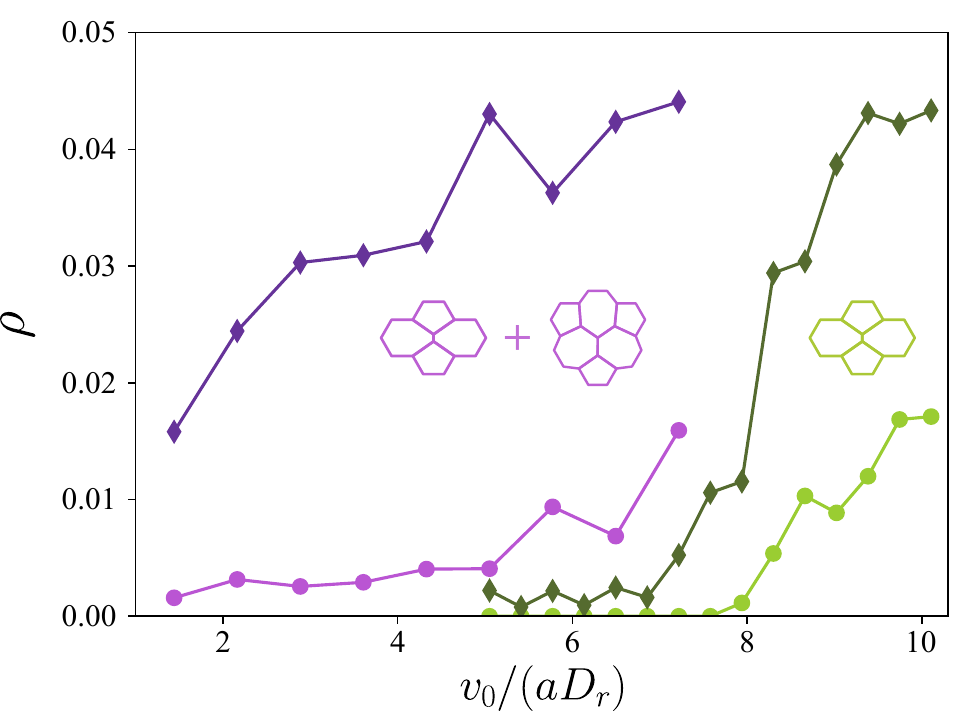}	
	\caption{Fraction of free dislocations (diamonds) and free disclinations (circles) in the total amount of cells, in the case of closed (green symbols) and open (magenta symbols) systems, in the multiphase field simulations, as a function of the self propulsion speed.}
\end{figure}